# Eye-Tracking Metrics for Task-Based Supervisory Control

Jeffrey R. Peters[1], Amit Surana[2], and Luca Bertuccelli[2]
[1]Center for Control, Dynamical Systems, and Computation, University of California, Santa Barbara
[2]United Technologies Research Center, East Hartford

*Task-based*, rather than *vehicle-based*, control architectures have been shown to provide superior performance in certain human supervisory control missions. These results motivate the need for the development of robust, reliable usability metrics to aid in creating interfaces for use in this domain.
To this end, we conduct a pilot usability study of a particular task-based supervisory control interface called the Research Environment for Supervisory Control of Heterogenous Unmanned Vehicles (RESCHU). In particular, we explore the use of eye-tracking metrics as an objective means of evaluating the RESCHU interface and providing guidance in improving usability. Our main goals for this study are to 1) better understand how eye-tracking can augment standard usability metrics, 2) formulate initial models of operator behavior, and 3) identify interesting areas of future research.

**Introduction**

*Eye-tracking and Supervisory Control*

The use of Unmanned Vehicles (UVs) is becoming commonplace in search and surveillance tasks (Nolin, 2012), and UVs often require feedback to varying degrees from one or several human operators. Due to the large amount of data that the operator is responsible for processing, operators are susceptible to *information overload*, which can cause degraded performance and potentially catastrophic consequences. Poor interface design can exacerbate these effects. Therefore, many researchers have begun to investigate human behavioral phenomena in their interactions with UV control interfaces, with the goal of better understanding the factors that affect human performance, modeling human behavior, and assessing system usability.

Eye-tracking has become a common tool for analyzing operator interactions with interfaces (Poole and Ball, 2006), and many researchers have found correlations between eye-tracking metrics and cognitive behavior. With this in mind, we seek to study eye-tracking metrics in the context of task-based human supervisory control. We believe these objective metrics can provide valuable insight into "typical use" for a given interface, which can, in turn, give insight into system usability and lead to better user interfaces.

Here, we present a pilot usability study in attempt to better understand human interaction with a particular supervisory control interface, and simultaneously explore the use of eye-tracking metrics as an objective means of augmenting subjective usability scores. Specifically, we use the task-based version of a widely studied interface called the Research Environment for Supervisory Control of Heterogenous Unmanned Vehicles (RESCHU) (Nehme, 2009). In our study, we collect objective gaze measurements while operators participate in a simulated control mission involving either 4 or 8 UVs. Through this data, we hope to better understand how eye-tracking can be used in modeling user behavior and evaluating the user interface. Finally, we use this insight to identify productive areas of future research.

*Related Literature*

RESCHU is widely studied in the context of human supervisory control (e.g., Nehme, 2009; Gartenberg, et al., 2013; Pickett, et al., 2013). The present work is largely in response to (Cummings, et al., 2014), in which researchers use RESCHU to conclude that task-based control architectures outperform vehicle-based control architectures in many respects. *Task-based* refers to a setup where the user does not have direct control over individual UV trajectories and can only issue high-level commands, whereas *vehicle-based* refers to a setup where the user can control individual UVs.

The standard means of evaluating user interfaces is through usability surveys. Some surveys, such as the *system usability scale* (SUS) (Brooke, 1996; Sauro and Lewis, 2012), have been studied extensively, and provide benchmark statistics for comparative purposes. These surveys usually provide general assessments of usability, rather than specific insight into how a particular interface could be improved.

Eye-tracking has been studied in a variety of contexts, particularly in psychology. One class of literature studies dynamic aspects of eye-movements, e.g. (Judd, et al., 2009). For our purposes, the most relevant works regarding eye-movements are those that are relevant to visual search (e.g., Zelinski, et al., 2008). Some such works show that eye-movements are highly dependent upon the task that is being performed (Borji and Itti, 2014). Other researchers seek to use eye-tracking metrics to quantify the mental state of the user, particularly workload (e.g., Recarte and Nunes, 2003). A few papers use eye-tracking metrics as feedback in the design of engineering systems (e.g., Bailey and Iqbal, 2008).

**Methods**

*The RESCHU Interface*

We start with a brief overview of task-based RESCHU. For a comprehensive treatment, the reader should consult (Cummings, et al., 2014; Nehme, 2009). RESCHU allows the user to control multiple UVs in a search and identification task. Fig. 1 shows a screen shot of the graphical user interface (GUI) and its 6 main features: (1) a map showing positions of UVs, targets, and hazard areas, (2) a search window to display the UV payloads (camera imagery),

(3) a message window to relay system information to the user, (4) a panel for engaging UV payloads, (5) a table showing arrival times for UVs to tasks, and (6) a panel for selecting damage tolerances and re-planning UV trajectories. The map depicts the UVs as bullet shapes according to MIL-STD-2525C convention, and depicts hazard areas and task locations as yellow circles and diamonds, respectively. The UVs incur damage when they intersect hazard areas on the map. Hazard areas stochastically appear and disappear, which creates a need for dynamic path planning.

Once a UV reaches a task location, an "ENGAGE" button on the panel (4) becomes active. When the operator presses the button, a surveillance image appears in the search window (2). Images are static, but can be panned/zoomed. A textual description of the search target appears on the message window (3). The operator then right-clicks the image where he/she believes the target to be. The UV is then automatically re-assigned to a new task and begins moving in a straight-line path to the new task location. The operator can change the UV flight paths in order to avoid damage. We focus on a task-based architecture (Cummings, et al., 2014). To this end, the operator is shown the current task assignments for all UVs in the map window (region (1) in Fig. 1). The operator can only change assignments and/or predicted flight paths by selecting a damage tolerance (low, medium, or high) in the re-plan panel (region (6)) and clicking the "Re-plan Times" button. The system then calculates the estimated arrival times for all vehicles. The operator can accept the new plan by clicking the check mark in the scheduling panel (region (5)) and the new arrival times are automatically tabulated.

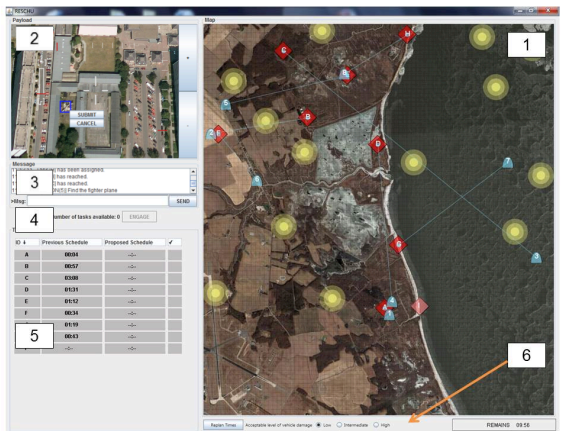

**Fig. 1:** RESCHU and its main features: (1) map window, (2) search window, (3) message window, (4) engage panel, (5) scheduling panel, and (6) re-plan panel.

*Participants*

We ran the test on 4 males (3 of which aged 18-34, one aged 35-44), 2 of which were given a mission with 4 UVs and 2 were given a mission with 8 UVs. All subjects indicated that their occupation was related to science and engineering. Due to experimental error we only collected complete eye-tracking data sets from 2 out of the 4 users, one of which performed a search task involving 4 UVs (Participant 3, age 18-34, no prior experience with RESCHU) and the other performed a task involving 8 UVs (Participant 4, aged 18-34, extensive experience with RESCHU, i.e., ``expert user"). 2 eye-tracking data sets is a small sample, but the objective of this pilot study is not to provide conclusive statistical insight, but rather to provide test data and preliminary intuition, which can give insight into usability and provide guidance to other researchers in designing future experiments.

*Procedure*

The setup consisted of a computer equipped with a Tobii x120 eye-tracking mechanism. The subject first completed a training module, and once he felt ready, the experimental investigator guided him through calibration of the eye-tracker and started the simulation.

The mission lasted 10 minutes, during which the participant interacted with RESCHU in the manner previously described. Participants had 2 main goals: 1) correctly process as many tasks as possible, i.e., find as many targets as possible, and 2) incur the least amount of UV damage. Time stamps for key events (engagement of search tasks, etc.) were logged by the RESCHU software. The eye-tracking mechanism did not make physical contact with the participants and did not impede their ability to interact with the interface. The eye-tracker collected data at a rate of 60 Hz, a sufficient sampling rate for analyzing macroscopic trends. Data that was logged by the eye tracker included a time stamp, horizontal and vertical positions of the subject's gaze on the screen (px), pupil diameter, and classifications of gaze events (fixations were defined using the built-in I-VT filter with the default threshold of 30 deg/s; for more information, refer to (User Manual, 2008).) The eye-tracker also sequentially numbered each gaze event and assigned a corresponding duration. Finally, a validation vector was included which assigned an integer between 0 (low uncertainty) and 4 (high uncertainty) according to the quality of each measurement.

Following the experiment, we gave each user a subjective usability questionnaire. The survey consisted of 3 parts. The first part contained 7 statements (3 positively worded, 4 negatively worded) for which users were asked to rate their degree of agreement on a Likert scale from 1-5. The statements dealt with general usability, and were a subset of the widely used SUS survey (Brooke, 1996). The second part contained multiple choice questions specific to RESCHU, and the last part asked for open comments.

**Results**

We refer to each occurrence of the user engaging a UV payload and subsequently searching for a target as a *search task*, and the corresponding times as *search times*. We quantify the length of one search task as the time between the user clicking "ENGAGE" and submitting the target location. We say that the user is engaged in a *nonsearch task* if they are not performing a search task, and we refer to corresponding times as *nonsearch times*. With respect to eye-tracking, we note that one can analyze many metrics including gaze events, gaze event durations, scan paths, static or dynamic pupil statistics, blink rates, saccade amplitude, peak velocities, and many more. In the results herein, we only present the following metrics, as we feel that these provide insight which is best aligned with the exploratory nature of this article and our goal of assessing system usability: gaze events in various regions of the GUI, fixation durations, fixation transition probabilities between regions, and scan paths.

*Subjective Measures*

We chose not to use the full SUS survey, but rather a subset, as we determined that 3 of the questions on the standard survey were not relevant to the task at hand. Therefore, to provide a quantitative analysis of system usability as a whole, we calculate a modified system usability score (MSUS) for each questionnaire based on user responses to the first section of the survey. This score is calculated analogous to the standard SUS score:

$$MSUS = \frac{25}{7}\left(17 + \left(\sum x_{P_i}\right) - \left(\sum x_{N_i}\right)\right),$$

where $x_{N_i}, x_{P_i}$ represent the scores for negatively and positively worded questions, respectively. The MSUSs for users 1- 4 based on their responses were 21.4, 71.4, 32.1, and 71.4, respectively, out of a maximum score of 100. Because of small sample size and large standard deviation, computing confidence intervals for these scores is not informative. However, one interesting point to note is that users for the 8 UV condition (users 2 and 4) had MSUSs that were significantly higher than that of the other users.

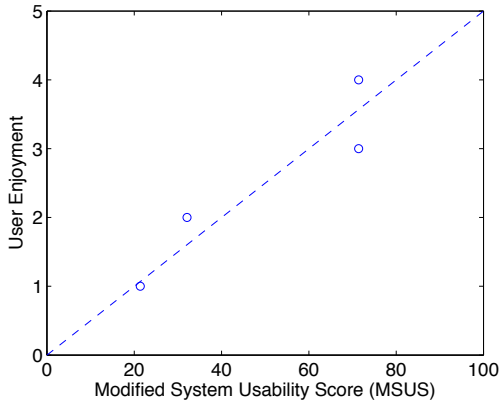

**Fig. 2: Subjective User Enjoyment vs. MSUS**

With regard to the second part of the survey, all users noted that the characters on the interface were too small. 2 users (users 1 and 3) indicated that they did not enjoy using the system, 1 user (user 4, the expert user) enjoyed using the system, and 1 user (user 2) had no opinion about system enjoyment. This question asked the users to rate their degree of enjoyment on a scale of 1-5. If we plot the user enjoyment versus the MSUS (Fig. 2) we see that the two measures align well, which serves to validate our assigned MSUSs. With respect the last portion of the survey, the most common remarks were that the characters were too small, the search window navigation was tedious, the search window size was too small, and the map window size was too big.

*Objective Measures*

In this section, we only consider participants for which we were able to obtain complete data sets (3,4). For participant 3, the eye-tracking mechanism logged 37,606 data points, 29,503 (78.3%) of which during search times. For participant 4, our expert user, the eye-tracking mechanism logged 37,088 data points, 26,087 (70.3%) of which during search times. Approximately 95% and 86% of the data points for participants 3 and 4, respectively, were accurate according to the validity vectors provided by the eye-tracker. Because gaze location varies slightly even during fixations, we filtered data by replacing raw gaze locations during fixation events with their average. Table 1 shows the percentage of gaze events during search, non-search, and total times, respectively, falling within each region (see Fig. 1). We note that the similarity in the proportion of participant 4's gaze that fell in the search window during search times (78.3%), and proportion of participant 3's eye-tracking data corresponding to search times (78.3%) is coincidental.

**Table 1: Distribution of Gaze Events**

*Participant 3*

|  | Search | Nonsearch | Total |
|---|---|---|---|
| Map Window | 4.2% | 50.9% | 16.4% |
| Search Window | 89.7% | 11.7% | 69.4% |
| Message Panel | 1.9% | 22.2% | 7.2% |
| Engage Panel | 0.0% | 4.3% | 1.4% |
| Re-Plan Panel | 0.1% | 4.9% | 1.4% |
| Unclassified | 4.0% | 5.9% | 4.5% |

*Participant 4*

|  | Search | Nonsearch | Total |
|---|---|---|---|
| Map Window | 1.0% | 43.9% | 13.7% |
| Search Window | 78.3% | 3.6% | 56.2% |
| Message Panel | 9.1% | 8.7% | 9.0% |
| Engage Panel | 3.2% | 6.5% | 4.3% |
| Re-Plan Panel | 0.4% | 3.6% | 1.3% |
| Unclassified | 6.7% | 9.8% | 7.6% |

For both participants, the percentage of gaze events falling within either the search or message windows was much higher during search times (participant 3: 91.6%, participant 4: 87.4%) than during nonsearch times (participant 3: 33.9%, participant 4: 12.3%). During nonsearch times, both users spent the most time looking in the map window (50.9%, 43.9%). If we examine combinations of windows with the most gaze events during nonsearch tasks, participant 3 spent most of his time looking within either the map window, search window, or the message window (84.8%), while participant 4 distributed his gaze mostly among the map window, the scheduling panel, and unclassified regions (81.1%). Neither of the participants looked at the re-plan panel very often.
To analyze scan-paths, i.e., sequences of fixations, we consider the fixations that occurred in the 6 major regions indicated in Fig. 1. Assuming that the sequence of fixations satisfies the 1-step Markov assumption (valid in certain circumstances (Towal, et al., 2013)) then we can construct a transition probability matrix by considering the probability that the next fixation will fall within a particular region, given the location of the current fixation. Fig. 3 and Fig. 4 show graphical representations of the transition probability matrices for participant 3 and participant 4 for 1) fixations during search tasks, and 2) fixations during nonsearch tasks. We separate the data in this way to highlight qualitative differences in the two conditions. In these plots, the six regions are represented by boxes, each with a vertex inside.

window and unclassified regions than did participant 3, especially during search times. However, they both show largely similar qualitative behaviors in all conditions. We note that we also analyzed differences in scan paths from a

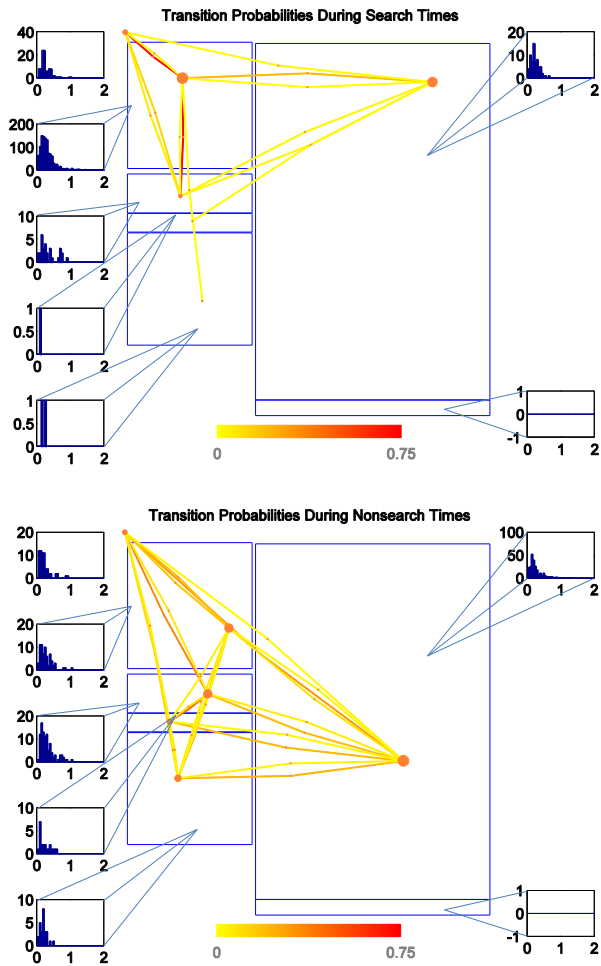

**Fig. 3: Fixation Transition Probabilities for Participant 3**

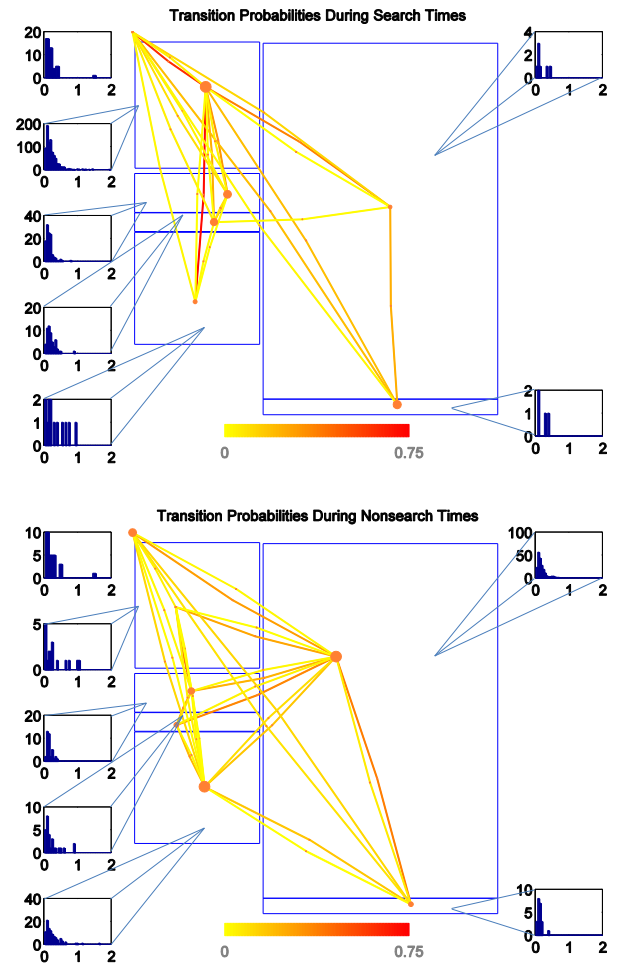

**Fig. 4: Fixation Transition Probabilities for Participant 4**

The vertex located outside the boxes represents unclassified regions. Lines connecting the vertices represent transitions between regions, with the direction of convex curvature representing a forward transition. The size of the vertex corresponds to the probability that the next fixation will fall in the same region as the current fixation. For clarity, if the total number of transitions starting from a given region was less than 5, these transitions were omitted from the diagrams. It is apparent that there are qualitative differences between scan-path behavior of the two conditions (search or nonsearch). Both participants showed increased probability of a self-loop in the search window, and a decreased probability of self-loops in the map window during search times. Further, the transition probability distribution is much more ``balanced'' during nonsearch times for both participants, i.e., there are much more equally distributed probabilities with respect to the transitions between regions. If we compare probabilities between the two participants, participant 4 showed a slightly higher tendency to transition to and from the scheduling quantitative point of view using the balanced Kullback-Leibler (KL) divergence as a metric between two Markov transition matrices. However, because our main discussion points are adequately illustrated through qualitative results, we omit an explicit discussion of quantitative measures. In addition to the Markov transition matrices, Figs. 3 and 4 contain histograms quantifying the length of the fixations in each region.

**Discussion**

*Augmenting Standard Usability with Eye-Tracking*

The MSUS provides a unique way of quantifying overall usability. Since we used a subset of the questions from the standard SUS questionnaire, conclusive comparisons between benchmark SUS scores and the MSUSs is difficult. Despite this, we note that the average SUS for a generic user interface is about 67 (Sauro and Lewis, 2012). From our data, 2 users (users 1,3) had MSUSs well below this benchmark, while the remaining 2 had MSUSs close to this benchmark.

The two users that had the highest MSUSs were the two users that had some degree of previous experience with RESCHU, and it is likely that their results were inflated by training. Therefore, we speculate that the mean MSUS for the RESCHU interface is most likely below the average SUS for a generic interface. Assuming the two measures are comparable in some way, this suggests RESCHU could be improved.

Augmenting subjective arguments with eye-tracking data can both corroborate data contained in subjective surveys and provide additional insight into specific improvements to the interface. At a basic level, the fact that our hardware did not produce valid data sets for two of the users already presents valuable insight. Indeed, the two users that did not produce valid data sets tended to squint their eyes and lean in toward the screen, preventing adequate data collection. This lack of data corroborates the survey comments, as both users indicated that the reason for this behavior was that text on the interface was too small, and that they had difficulty seeing certain portions of the screen (particularly the search window). On the other hand, the eye-tracking data that was collected, particularly that in Table 1, Fig. 3, and Fig. 4 can potentially provide more specific insight. Since both users showed similar trends in the amount of time they spent looking in the various regions, it is conceivable that re-sizing the various windows to be more consistent with these patterns could improve performance. For example, it may be advantageous to make the scheduling window smaller, since it currently takes up a large portion of the screen, but neither user focused on this portion very often. Differences that are present between the search and nonsearch conditions in the gaze-patterns in Fig. 3 and Fig. 4 also indicate that using a dynamic re-sizing scheme might improve performance. In this case, our data suggests that blowing up the search window during searches and shrinking other windows, especially the scheduling panel and the map window, could have beneficial effects.

*Formulating Initial User Models*

Since both users qualitatively showed largely similar patterns in their gaze transition probabilities, Figs. 3 and 4 allow us to postulate about "typical" use of the system. During nonsearch tasks, both users mostly looked at the map window, but also spent significant time shifting their gaze among the different regions. During search times, users fixated mostly on the search window, with very few gaze transitions. When the users did shift their gaze out of the search window during a search task, they generally transitioned back quickly. This suggests that typical use may involve a balanced gaze approach in which the user constantly scans the screen (with slight bias toward the map window) during nonsearch times, and an unbalanced gaze approach during search times, where the user spends almost all of their time in the search window with occasional short looks outside. Despite similarities between users, there are still differences in gaze behaviors that could potentially be exploited by system designers. For example, User 4 (the expert user) seemed to have a slightly more balanced approach during search times, then did user 3. Of course, this could be a result of individual differences, but it could also be a result of increased workload, since increased tendency to transition outside the search window during search times could be due to pressure to attend to other UVs.

The data contained in Figs. 3 and 4 also has the potential to produce a multi-layered user model. Indeed, at the highest level the user can be modeled as a state machine, with states such as "search" and "nonsearch", while within each state we can model user behavior on a finer level using Markov relations similar to those contained in Fig. 3 and 4. The utility of such a model should be further assessed.

*Areas of Future Research*

Eye-tracking shows potential as a means of augmenting standard system usability metrics and modeling user behavior in task-based supervisory control environments. Further research is necessary to determine if observed trends are significant, or due to other mechanisms. We have postulated that the study of gaze data, particularly scan paths and the proportion of gaze events falling in various regions of the interface, could provide insight into specific areas of improvement. With regard to RESCHU, it is of interest to investigate whether resizing windows according to gaze event proportions or implementing dynamic resizing schemes does in fact produce improved perceived system usability. Future research should also provide a thorough treatment of other metrics that eye-tracking can provide and investigate its relation to usability. With regard to user models, further investigation is needed to see if our postulated behavior is observed for general users, and if so, what changes to this pattern emerge as we alter the GUI. Our data also suggests that scan-paths might be an indicator of operator workload. As such, it would be advantageous to further explore the Markov assumption, and formulate more explicit representations of scan-path behavior. Finally, incorporating fixation times on each UV into visual scan-path models may also be useful. More thorough analysis of user abstractions derived using rigorous Expectation-Maximization models in conjunction with scan-path models should be investigated as well.